\documentclass[a4paper]{jpconf}
\usepackage{amssymb}
\usepackage{graphicx}

\newcommand{\be}{\begin{equation}}
\newcommand{\ee}{\end{equation}}
\newcommand{\bra}{\langle}
\newcommand{\ket}{\rangle}
\newcommand{\bea}{\begin{eqnarray}}
\newcommand{\eea}{\end{eqnarray}}
\newcommand{\dis}{\displaystyle}

\begin{document}
\title{Multiple Time Series Ising Model for Financial Market Simulations}

\author{Tetsuya Takaishi}

\address{Hiroshima University of Economics, Hiroshima 731-0192, JAPAN}

\ead{tt-taka@hue.ac.jp}

\begin{abstract}
In this paper we propose an Ising model which simulates multiple financial time series. 
Our model introduces the interaction which couples to spins of other systems. 
Simulations from our model show that time series exhibit the volatility clustering that is often observed in the real financial markets. 
Furthermore we also find non-zero cross correlations between the volatilities from our model.  
Thus our model can simulate stock markets where volatilities of stocks are mutually correlated.
\end{abstract}

\vspace{-5mm}
\section{Introduction}
The financial markets are considered to be complex systems where many agents are interacting at different levels
and acting rationally or in some cases irrationally.
Such financial markets produce a rich structure on time variation of various financial assets
and the pronounced properties of asset returns has been classified as the stylized facts, e.g. see \cite{CONT}.
The most prominent property in the stylized facts is that asset returns show 
fat-tailed distributions that can not be explained by the standard random work model.
A possible explanation for the fat-tailed distributions is 
that the return distributions are viewed as a finite-variance mixture of normal distributions, 
suggested by Clark\cite{Clark}.
In this view asset returns follow a Gaussian random process with a time-varying volatility. 
This view, using realized volatility\cite{RV1,RV2} constructed from high-frequency data, 
has been tested and 
it is found that the asset returns  are consistent with this view\cite{Andersen1,Andersen2,Andersen3,Andersen4,TakaishiRV,TakaishiRV2}.

Bornholdt proposed an Ising model designed to simulate financial market as a minimalistic agent based model\cite{Bornholdt}.
The Bornholdt model successfully exhibits several stylized facts such as fat-tailed return distributions 
and volatility clustering\cite{Bornholdt,Yamano,Bornholdt2,Bornholdt3}. 
Variants of the model have also been proposed 
and they exhibit exponential fat-tail distributions\cite{Potts} or asymmetric volatility\cite{Asymmetric}. 
The view of finite-variance mixture of normal distributions  has also been tested for the Bornholdt models and
it is shown that returns simulated from the models are consistent with 
the  view of the finite-variance mixture of normal distributions\cite{RVspin1,RVspin2}.
 
So far the models studied  only dealt with a financial market where a single asset is traded.
In the real financial markets various financial assets are traded and they are correlated each other. 
Measuring correlations between assets is important to investigate stability of financial markets 
and many studies have been conducted to reveal properties of correlations in financial markets, e.g. \cite{CC0,CC1,CC2,CC3,CC4}.
In this study we propose an Ising  model that extends the single Bornholdt model to a multiple time series model.
Then we perform simulations of our model and show that the model can exhibit correlations between the return volatilities.

\section{Multiple Time Series Ising Model}
Let us consider a financial market where $K$ stocks are traded and
assume that each stock is traded by $N=L\times L$ agents on a square lattice. 
Each agent has a spin $s_i$ and 
it takes two states $s_i=\pm 1$ corresponding to "buy" (+1) or "sell" (-1), where $i$ stands for the $i$th agent.
The decision of agents are made probabilistically according to a local field.
In our model the local field $h_i^{(k)}(t)$ at time $t$ is given by
\be
h_i^{(k)}(t)=\sum_{\bra i,j\ket}J s_j^{(k)}(t) -\alpha s_i^{(k)}(t)|M^{(k)}(t)|  + \sum_{j=1}^K \gamma_{jk}M^{(j)}(t),
\label{eq:h}
\ee 
where $\bra i,j\ket$ stands for a summation over the nearest neighbor pairs, $k$ denotes the $k$th stock 
and $M^{(k)}(t)$ is the magnetization that shows an imbalance between "buy" and "sell" states, given by 
$M^{(k)}(t)=\sum_{l=1}^N s_l^{(k)}(t)$. $J$ is the nearest neighbor coupling and in this study we set $J=1$. 
The third term on the right hand side of eq.(\ref{eq:h}) describes the interaction with other stocks 
that is not present in the Bornholdt model.
More precisely this interaction couples to the magnetization of other stocks 
and introduce an effect of imitating the states of other stocks.
The magnitude of the interaction is given by the interaction parameters that form a matrix $\gamma_{lm}$ 
having zero diagonal elements, i.e. $\gamma_{ll}=0$.
As in the Bornholdt model 
the states of spins are updated according to the following probability.
\bea
\label{eq:Prob}
s_i^{(k)}(t+1) & =   +1  &  p=1/(1+\exp(-2\beta h_i^{(k)}(t))), \\ \nonumber
s_i^{(k)}(t+1) & =  -1    & 1- p.
\eea

\begin{figure}[h]
\vspace{5mm}
\begin{minipage}{0.5\hsize}
\begin{center}
\includegraphics[height=5cm]{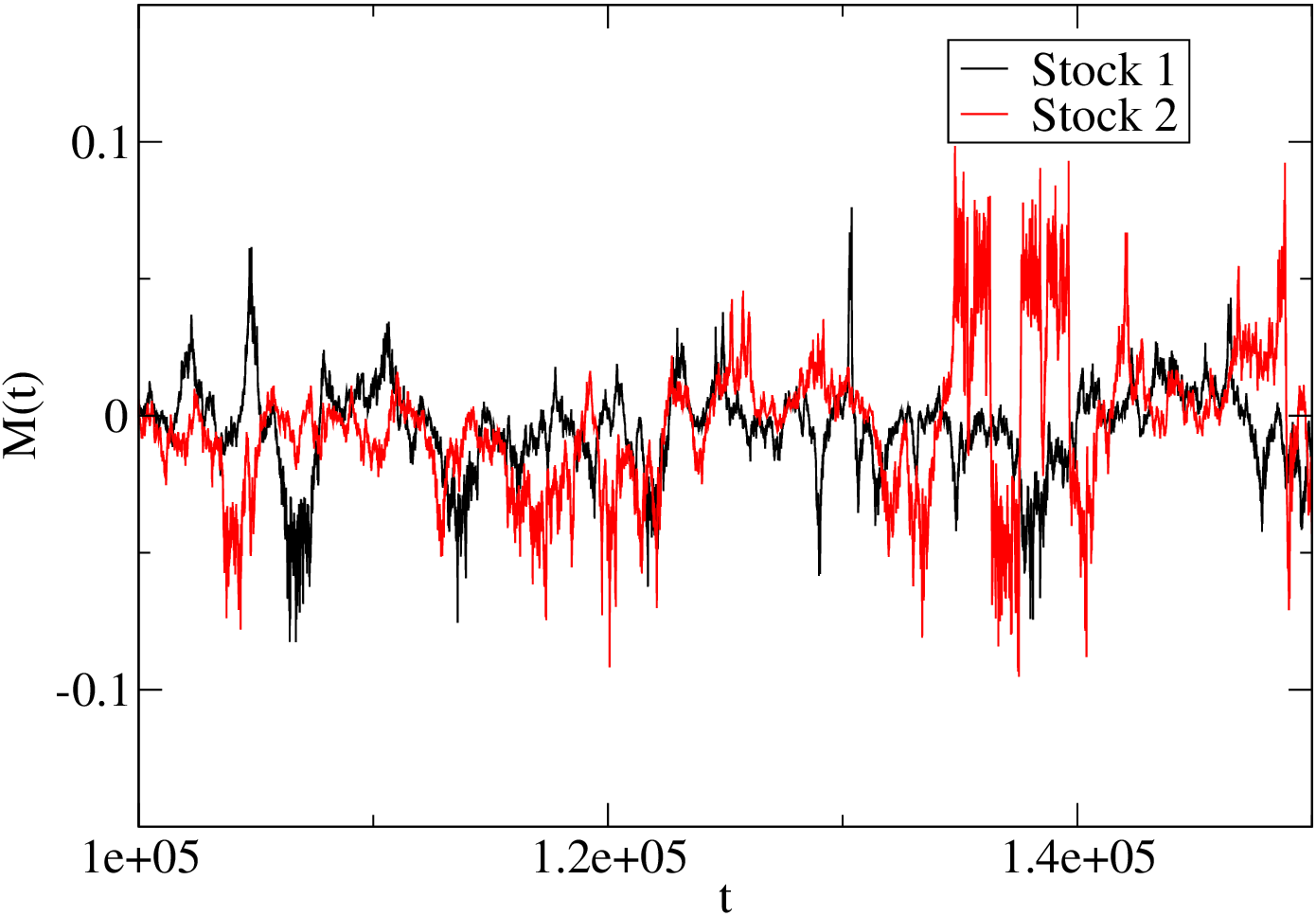}
\caption{Time series of the magnetization for stocks 1 and 2 at $\gamma=0.0$.
}
\label{fig:dH}
\end{center}
\end{minipage}
\hspace{3mm}
\begin{minipage}{0.5\hsize}
\begin{center}
\centering
\vspace{1mm}
\includegraphics[height=5cm,keepaspectratio=true]{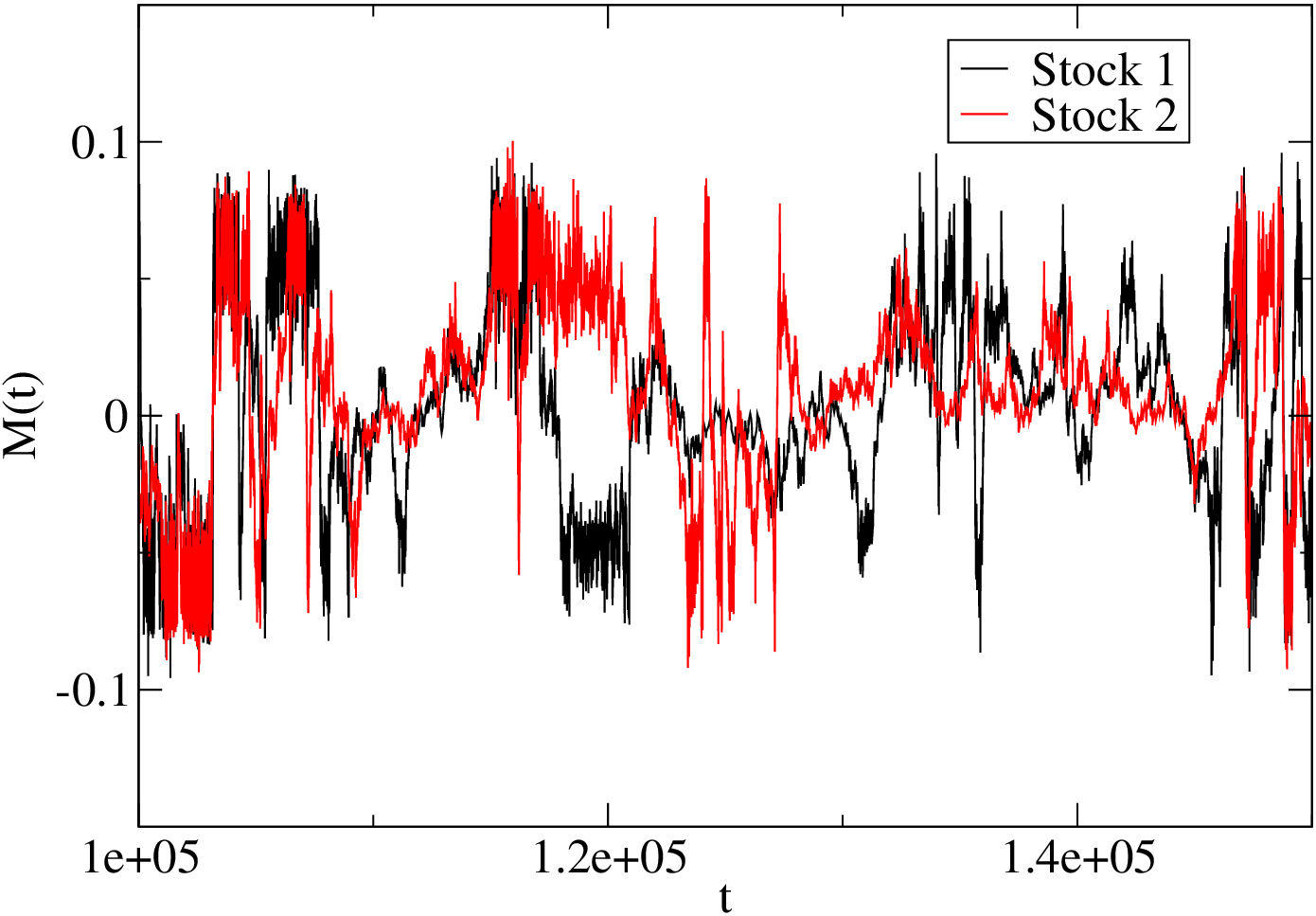}
\caption{Time series of the magnetization for stocks 1 and 2 at $\gamma=0.15$.
}
\label{fig:Acc}
\end{center}
\end{minipage}
\end{figure}

\section{Simulations}
In this study we perform simulations for $K=2$, i.e. we simulate a stock market consisting of two stocks. 
The simulations are done on  $120 \times 120$ square lattices with the periodic boundary condition.
We set simulation parameters to $(\beta,\alpha,J)=(2.0,30,1)$.
Here we assume symmetric $\gamma$, i.e. $\gamma_{12}=\gamma_{21}$, 
which means that the stocks we consider give the same interaction effect to other stocks each other. 
Here we make simulations for five values of $\gamma$, $\gamma=(0.0,0.05,0.07,0.10,0.15)$.
The states of spins are updated randomly according to eq.(\ref{eq:Prob}). 
After discarding the first $10^4$ updates as thermalization we collect data from $5\times 10^5$ updates.

\begin{figure}[h]
\vspace{-10mm}
\begin{center}
\includegraphics[height=8.3cm,width=15cm]{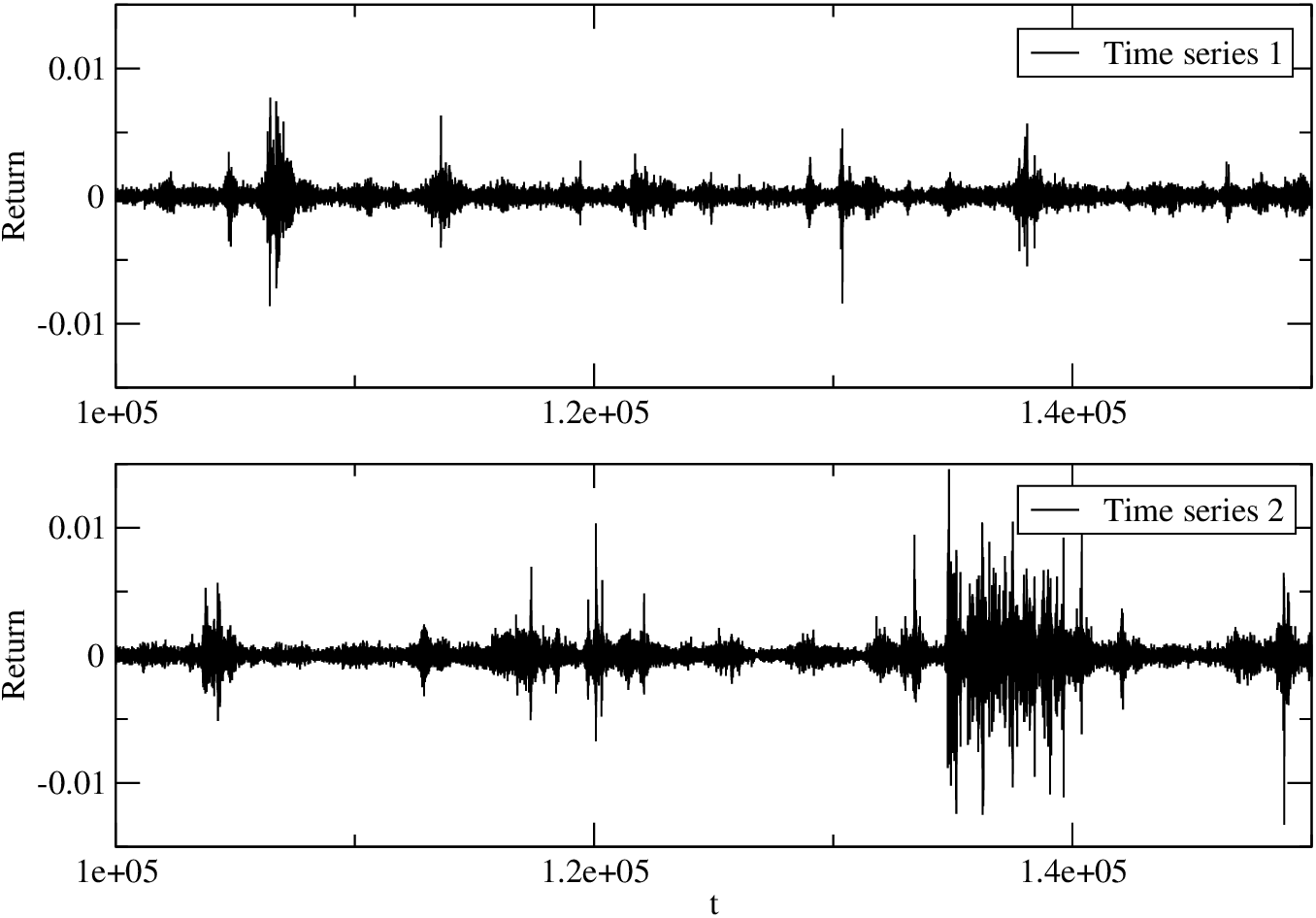}
\end{center}
\vspace{-5mm}
\caption{
Time series of return for stocks 1 and 2 at $\gamma=0.0$.
}
\end{figure}

\begin{figure}[h]
\vspace{10mm}
\begin{center}
\includegraphics[height=8.3cm,width=15cm]{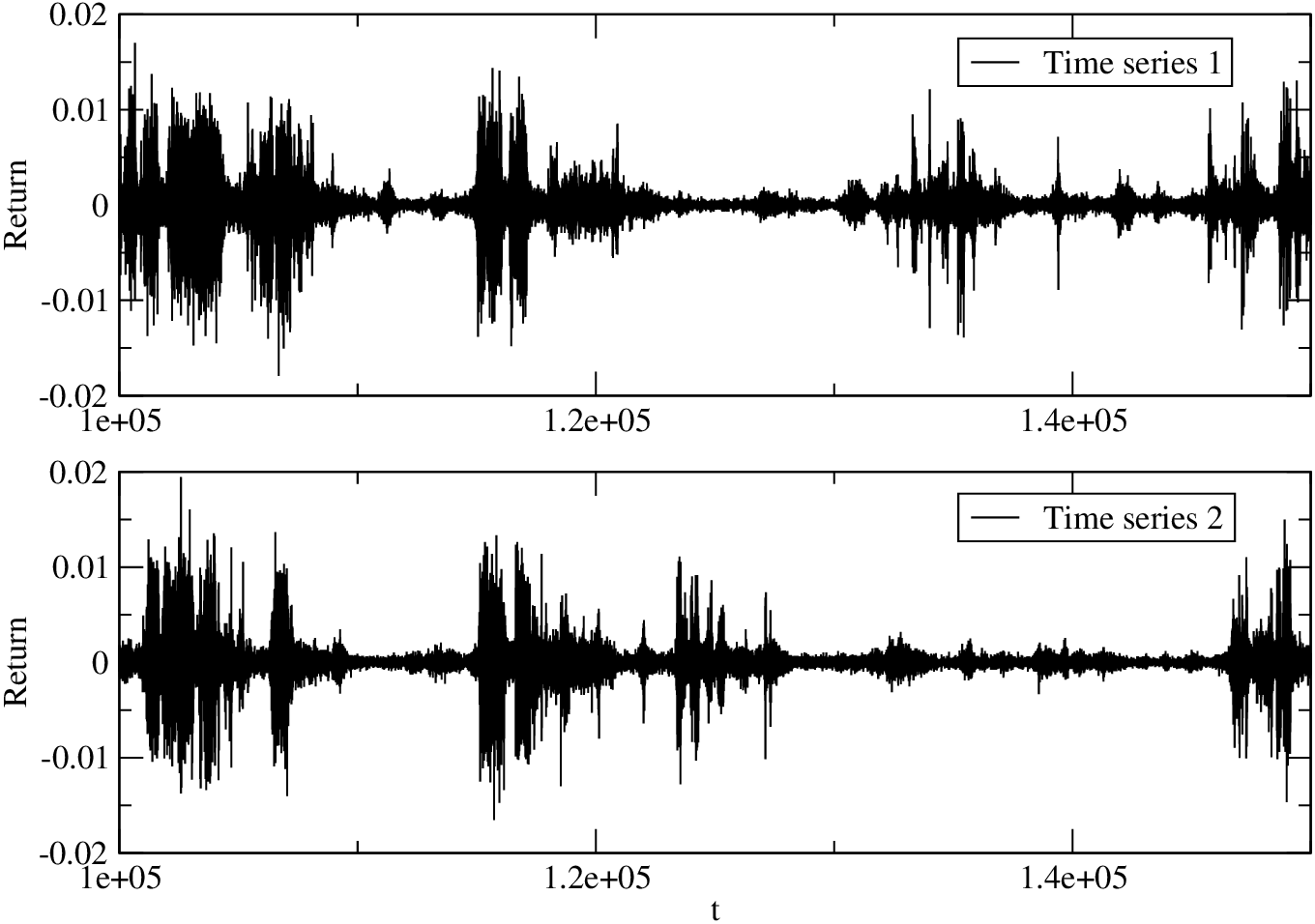}
\end{center}
\vspace{-5mm}
\caption{
Time series of return for stocks 1 and 2 at $\gamma=0.15$.
}
\end{figure}

Fig.1 shows the time series of the magnetization for two stocks ( 1 and 2 ) at $\gamma=0.0$. 
Simulating at $\gamma=0.0$ means that two stocks are independent.
On the other hand  Fig.2 shows the time series of the magnetization at $\gamma=0.15$
where the interaction between stocks are present and
we find that synchronization occurs between the magnetizations.
For other $\gamma>0$ we also see similar synchronization between the magnetizations to a certain extent.

Such synchronization also occurs between the returns defined by
$Return(t)=(M(t+1)-M(t))/2$ as in \cite{Bornholdt2}.
The time series of returns at $\gamma=0.0$ and 0.15 are shown in Figs. 3 and 4 respectively.
It is clearly seen that volatility clustering occurs in the return time series.
At $\gamma=0.0$, however, we see no synchronization between the volatility clusterings.
On the other hand some volatility clusterings synchronize at $\gamma=0.15$.

To quantify the strength of the synchronization 
we measure cross correlations between volatilities from the two stocks.
The cross correlation between the $l$th and the $m$th volatilities is given by 
$\dis \frac{\bra (v^{(l)}(t)-\bra v^{(l)}(t)\ket)(v^{(m)}(t)-\bra v^{(m)}(t)\ket) \ket }{\sigma^{(l)}\sigma^{(m)}}$,
where $v^{(l)}$ and $\sigma^{(l)}$ stands for the volatility and its standard deviation of the $l$th stock respectively.
Here we define the volatility by the absolute value of return.
Table 1 shows the values of cross correlations for various $\gamma$ and
we find that the cross correlation of volatility increases with $\gamma$. 

\begin{table}[t]
\vspace{-2mm}
\centering
\caption{Cross correlation between the volatilities for various $\gamma$.}
\begin{tabular}{l|ccccc}
\hline
$\gamma$  & 0.0 & 0.05 & 0.07 & 0.10 & 0.15 \\
\hline
cross correlation &  $8.2\times 10^{-3}$ & $5.8\times 10^{-2}$ &  $8.4\times 10^{-2}$ &  0.15 & 0.31\\
\hline
\end{tabular}
\vspace{-2mm}
\end{table}

\section{Conclusion}
We have proposed an Ising model which can simulate multiple financial time series
and performed simulations of a stock market with two stocks. 
The interaction parameter $\gamma$ in the model tunes the strength of correlations between stocks.
We calculated cross correlation between volatilities of the two stocks and
found that the cross correlation increases with $\gamma$. 
Therefore our model serves to simulate stock markets where volatilities of stocks are mutually correlated.
Since we have demonstrated the model including only two stocks
it may be desirable to further investigate the more realistic model that includes 
many stocks.

\section*{Acknowledgement}
Numerical calculations in this work were carried out at the
Yukawa Institute Computer Facility
and the facilities of the Institute of Statistical Mathematics.
This work was supported by JSPS KAKENHI Grant Number 25330047.


\section*{References}
\begin{thebibliography}{9}

\bibitem{CONT}
Cont R 2001
{\it Quantitative Finance} {\bf 1}  223--236

\bibitem{Clark}
Clark  P K 1973
{\it Econometrica} {\bf 41}  135-155


\bibitem{RV1}
Andersen T G and Bollerslev T 1998
\textit{International Economic Review} {\bf 39}  885-905

\bibitem{RV2}
Andersen T G,  Bollerslev T, Diebold F X and Labys P 2001
\textit{J. Am. Statist. Assoc.}A {\bf 96}  42-55


\bibitem{Andersen1}
Andersen T G,  Bollerslev T,  Diebold F X and Labys P 2000
\textit{Multinational Finance Journal} {\bf 4}  159--179


\bibitem{Andersen2}
Andersen T G,  Bollerslev T,  Diebold F X and Ebens H 2001
\textit{Journal of Financial Economics} {\bf 61}  43--76


\bibitem{Andersen3}
Andersen T G,  Bollerslev T and Dobrev, D 2007
{\it Journal of Econometrics}  {\bf 138}  125-180

\bibitem{Andersen4}
Andersen T G,   Bollerslev T, Frederiksen P and Nielsen M {\O} 2010
{\it Journal of Applied Econometrics} {\bf 25}  233-261


\bibitem{TakaishiRV}
Takaishi T, Chen TT and Zheng Z 2012
\textit{Prog. Theor. Phys. Supplement} {\bf 194}  43-54

\bibitem{TakaishiRV2}
Takaishi T 2012
{\it Procedia - Social and Behavioral Sciences} {\bf 65} 968--973


\bibitem{Bornholdt}
Bornholdt S 2001
{\it Int. J. Mod. Phys. C} {\bf 12}  667--674

\bibitem{Yamano}
Yamano Y 2002 
\textit{Int. J. Mod. Phys. C} {\bf 13} 89-96

\bibitem{Bornholdt2}
Kaizoji T, Bornholdt S and Fujiwara Y 2002
{\it Physica A} {\bf 316}  441--452

\bibitem{Bornholdt3}
Krause S M and Bornholdt S 2011
{\it arXiv:1103.5345}


\bibitem{Potts}
Takaishi T 2005
{\it Int. J. Mod. Phys. C} {\bf 16}  1311--1317

\bibitem{Asymmetric}
Yamamoto R 2010 
\textit{Physica A} {\bf 389} 1208-214

\bibitem{RVspin1}
Takaishi T 2013
\textit{J. Phys.: Conf. Ser.} {\bf 454} 012041 

\bibitem{RVspin2}
Takaishi T 2014
\textit{JPS Conf. Proc.}  019007 

\bibitem{CC0}
Plerou V, Gopikrishnan P, Rosenow B, Amaral L A N and Stanley H E 1999
\textit{Phys. Rev. Lett. }{\bf 83} 1471-1474

\bibitem{CC1}
Plerou V,  Gopikrishnan P, Rosenow B, Amaral L A N,  Guhr T and Stanley H E  2002
\textit{Phys. Rev. E }{\bf 65} 066126

\bibitem{CC2}
Utsugi A, Ino K and Oshikawa M 2004
\textit{Phys. Rev. E }{\bf 70} 026110

\bibitem{CC3}
Kim D and Jeong H 2005
\textit{Phys. Rev. E }{\bf 72} 046133

\bibitem{CC4}
Wang D, Podobnik B, Horvati\ifmmode \acute{c}\else \'{c}\fi{} D and Stanley H E 2011 
\textit{Phys. Rev. E }{\bf 83} 046121


\end{thebibliography}
\end{document}